\documentstyle[12pt]{article}

\newcommand{\be}{\begin{equation}}
\newcommand{\ee}{\end{equation}}
\newcommand{\bea}{\begin{eqnarray}}
\newcommand{\eea}{\end{eqnarray}}

\newcommand{\del}{\partial}

\newcommand{\bs}{|B\rangle}
\newcommand{\tq}{\tilde{q}}

\newcommand{\nbox}{{\,\lower0.9pt\vbox{\hrule \hbox{\vrule height 0.2 cm \hskip
0.2 cm \vrule height 0.2 cm}\hrule}\,}}

\begin{document}
\begin{titlepage}
\title{ 
        \begin{flushright}
        \begin{small}
        RUNHETC-99-32\\
        hep-th/9909017\\
        \end{small}
        \end{flushright}
        \vspace{1.cm}
 D-Branes in Linear Dilaton Backgrounds}

\author{
Arvind Rajaraman\thanks{e-mail: \tt arvindra@physics.rutgers.edu}
\ and
Moshe Rozali\thanks{e-mail: \tt rozali@physics.rutgers.edu}\\
\\
        \small\it Department of Physics and Astronomy\\
        \small\it Rutgers University\\
        \small\it Piscataway, NJ 08855
}

\maketitle

\begin{abstract}
 We construct a Dirichlet boundary state for linear dilaton backgrounds. The 
state is conformally invariant and satisfies Cardy's conditions. We apply
this construction to two dimensional  string theory.

\end{abstract}

\end{titlepage}

\section{Introduction}

  The study of Dirichlet branes \cite{pol} has played an essential role in 
the understanding of non-perturbative physics of various string vacua.
 In the strong coupling regime the D-branes become the lightest
BPS states, and therefore play the role of the fundamental degrees of 
freedom in a dual formulation. D-instantons, similarly, are important in
giving non-perturbative contributions to scattering amplitudes.

Given this success, one is motivated to study D-branes in string vacua 
with less symmetries, and sometimes with no clear geometrical interpretation 
\cite{dbranes}. Some of these string vacua are holographically dual (in 
the sense of \cite{mald}) to
non-gravitational theories, and the D-brane
 spectrum and interactions 
are important in the dynamics of those theories.

 The study of D-brane in general backgrounds is difficult, and a 
general method of construction is unknown. We review the boundary state 
approach to the subject in section 2. In particular we review a 
construction of boundary states due to Cohen, Moore, Nelson and  Polchinski
\cite{cmnp}. This construction can be useful in conformal field theories
where the spectrum  has a simple structure. 

 The examples studied in this paper are conformal field theories with 
a linear dilaton component. We review such backgrounds in section 3, and 
construct a Dirichlet boundary state for such backgrounds. (We note that
 Dirichlet boundary states in linear dilaton background
have also  been considered in \cite{Li}. 
 The boundary state constructed there appears to be
 conformally invariant, but does
 not satisfy Cardy's  conditions, defined below.)

 In section 4 we demonstrate the construction by explicitly constructing
D0-branes and D-instantons for the case of 2d string theory \cite{Liouville}.
 Those objects are relevant in studying non-perturbative questions in 
2d string theory.  We describe 
 further applications and directions for future research in the final section.
An appendix contains a generalization of the boundary state to incorporate 
worldsheet supersymmetry.

\section{General D-branes}

The general construction of a D-brane in an arbitrary string
background is not known.
In this section we outline an approach to this problem
based on \cite{cmnp}.
In the
process of doing so, we review some well-known
facts about open string theories, and introduce our
notation.

To make the presentation simpler, we concentrate on the
matter
 part of the CFT. One may choose
to work in light-cone variables as in \cite{gutp}, or add a
ghost sector and perform a BRST quantization \cite{boundary}.
In a conventional string background, where the
ghosts decouple from the matter CFT, their treatment
does not depend on the particular
background chosen.

The simplest way to construct an open string
theory is
to impose boundary conditions on the string variables and perform
a canonical quantization in the open string
sector.
The procedure requires a Lagrangian description
of the CFT, and is hard to implement in general string
backgrounds.

We instead work in the boundary state formalism \cite{boundary}.
In this formalism, we are given a `bulk' CFT, that
is,  a closed string background. This bulk CFT is
specified by a set of bulk operators and
their OPEs. These OPE coefficients are required to satisfy
nonlinear constraints: associativity and modular invariance on
the torus. These conditions guarantee consistency of
the theory formulated on an arbitrary Riemann surface.

Given the bulk CFT, we define a D-object to be an
open string theory consistent with the bulk CFT.
The building blocks of an open CFT, and the conditions
they satisfy, were analysed in \cite{sew}. A complete
specification of
an open string theory involves the following elements:

-A boundary state $|B\rangle$ which is a generalized coherent state in
the closed string Hilbert space. To
preserve the conformal invariance, essential for having 
a consistent string theory, one requires:
\be
\label{conf}
L_n-\tilde{L}_{-n}\bs =0
\ee

The state $\bs$ specifies the tree level couplings of the
D-object to all closed string fields.

- A set of open string operators (also called boundary 
operators, or more precisely ``boundary condition
preserving boundary operators''.)

- Closed string OPEs, open string OPEs, and open-closed OPEs.

These building blocks have to satisfy nonlinear relations \cite{sew}, the 
sewing constraints. These conditions guarantee the
consistency of the theory on an arbitrary (oriented) Riemann surface
with boundaries \cite{sew}. 

In the following, we study boundary states in linear dilaton
backgrounds. We emphasize that in principle, the
boundary state alone does not specify the open string theory.
One has to supply an open string operator algebra which satisfies
the sewing constraints. The existence of such an operator
algebra might impose further conditions on the possible
boundary states. Furthermore, such an operator algebra might
not be unique, in which case distinct D-objects share the same
boundary state.

The unique constraint involving only the boundary state is
Cardy's condition \cite{Cardy}, introduced below.

Our starting point is a solution of the conformal invariance
condition (\ref{conf}) based on any Virasoro module in the theory.
Suppose we are given a level matched
primary of the Virasoro algebra of dimensions $h_L=h_R=h$. Define
\be
\label{contra}
|B\rangle_h = \sum_{I,J} M_{IJ}^{-1}L_{-I}\tilde{L}_{-J}|h\rangle
\ee
Here $I,J$ are ordered strings of indices $n_1\cdots n_r$, and
\be
L_I=L_{n_1}\cdots L_{n_r}.
\ee
The contravariant form is defined as
\be
M_{IJ}=\langle h|L_IL_{-J}|h\rangle=\langle h|\tilde{L}_I\tilde{L}_{-J}|h\rangle
\ee

$M_{IJ}$ is invertible for any Virasoro module. For degenerate
modules, one has to mod out by the null vectors.

It is easy to see that $|B\rangle_h$ is conformally invariant by
showing
that $(L_n-\tilde{L}_{-n})|B\rangle_h$  is orthogonal to 
all states in the module based on
$|h\rangle$. Furthermore, by an application of Schur's lemma, the
solution is unique in each Virasoro module \cite{pvt,also}. A similar 
construction can be given
for any chiral algebra, yielding a boundary state preserving
one copy of that chiral algebra \cite{Ish}. In particular, we 
extend the discussion to include worldsheet supersymmetry in appendix A.

The boundary states $|B\rangle_h$ are the building blocks of any
physical boundary state $\bs$. The physical boundary state is required
to satisfy Cardy's conditions, which guarantee the existence of 
open string quantization of the system. We regard this condition as
 an effective way of
finding a basis of the physical boundary states
in the theory. 

  Cardy's conditions on the boundary state are obtained as follows. The
partition function on the annulus can be computed in the closed
string sector as
\bea
\label{annulus}
Z_B=\langle B| e^{-2\pi l (L_0+\tilde{L}_0-{c\over 12})}\bs
=\langle B|\tilde{q}^{(L_0+\tilde{L}_0-{c\over 12})}\bs
\eea
where $l$ is the closed string modulus of the annulus, and
$\tilde{q}= e^{-2\pi l}$.

This partition function can be written in terms of
the open string modulus, $q=e^{-\pi t}$ (with $t={1\over 2l}$
being the open string modulus). As a power series in $q$, one can read
from $Z_B$ the dimensions and multiplicities of
the open string operators. As such, all coefficients in the series must be 
nonnegative integers. The coefficient of the unit operator must be one 
(up to an overall
multiplicative factor in the boundary state, interpreted as the
number of D-branes.)

Finally, in the case of noncompact bosons, one
has to allow a slight generalization. Associated with
a noncompact boson one has an integration over
 continuous  momentum in either the open string sector
or the closed string sector.
Such integrations lead to factors which are 
nonanalytic in either $q$ or $\tq$. We 
generalize Cardy's conditions to
allow such factors, since their origin
is clear for noncompact bosons. In fact such factors
can be helpful in interpreting the boundary
state. For a free boson, a Dirichlet boundary
state allows an arbitrary momentum
in the closed string sector, while in the Neumann boundary
state the open strings can carry arbitrary
momentum. We generalize this to
any noncompact boson, by
defining a ``Dirichlet'' boundary state as having no free momentum
for the open strings, corresponding intuitively  to
open strings with fixed endpoints.

\section{Linear Dilaton Backgrounds}

Conformal field theories with a linear dilaton  appear as an 
ingredient in many
string backgrounds, critical and non-critical. 
For example, they appear in the NS 5-brane theory \cite{chs},they 
provide an explicit Lagrangian formulation
of Virasoro minimal models \cite{cft} and WZW models \cite{cft}, and
are an essential part of 2d string theory \cite{Liouville}.
The
construction of the Dirichlet boundary state
should apply to all such backgrounds.

We demonstrate the construction for the case of the
2d string theory. The worldsheet theory is
Liouville theory coupled to a single boson  $X$.
The worldsheet action is then 
\be
{\cal L}=\del X \bar{\del} {X} +\del \phi \bar{\del} {\phi}+
\mu e^{\gamma\phi}+ Q\phi  R
\ee
with $Q=2\sqrt{2}, \gamma=\sqrt{2}$ chosen to define a critical string theory.

The treatment of the free boson is standard, and we
focus on the field $\phi$.
The states in Liouville theory form a continuum  $|p\rangle$, $p\geq 0$ 
with dimensions $h_p={1\over 2}p^2+{1\over 8}Q^2$.
In addition, there is a discrete set of states, the special
states, which form a set of measure zero.
The states above do not correspond to  local operators, and
as normalizable modes are the appropriate ones to form a coherent
state in the closed string Hilbert space.

For generic $p$, the Virasoro module based on $|p\rangle$ is
nondegenerate. So it is straightforward to compute the
annulus diagram for the boundary state $|B\rangle_p$
(defined as in eq.(\ref{annulus}))
\be
{_p}\langle B|e^{-2\pi l (L_0+\tilde{L}_0-{c\over 12})}|B\rangle_p
= \tilde{q}^{p^2}\prod_n (1-\tilde{q}^{2n})\tilde{q}^{-{1\over 24}}
\ee
The result is identical to the corresponding
calculation for a single free boson.
One can construct then
\be
\label{bost}
\bs_D\equiv \sum_{p} e^{-ipX_0}|B\rangle_p.
\ee
 To check Cardy's conditions we compute the annulus diagram with the boundary
state $\bs_D$. After modular transformation, we find
\be
{_D}\langle B|e^{-2\pi l (L_0+\tilde{L}_0-{c\over 12})}\bs_D
=\prod_n (1-{q}^{2n}){q}^{-{1\over 24}}
\ee
where $q$ is the open string modulus defined above.

Therefore the boundary state  $\bs_D$ satisfies Cardy's 
conditions. There are no
logarithmic factors in the open string channel, as appropriate
for a Dirichlet boundary state.

We note that the above boundary state has a similar form  to the boundary
state
of a single free boson. This resemblence is misleading. The Liouville theory
is not exactly solvable, and a complete specification
of the open string theory would be different from that of D-branes
in flat space. However, the boundary state itself
is sensitive only to the spectrum, therefore it has a simple form when 
expressed in terms of  Virasoro modules.
 When expanded in modes of $\phi$, the 
state looks very different from the usual Dirichlet boundary state
 in flat space.

We mention in passing that Neumann boundary conditions in linear
dilaton backgrounds were discussed in \cite{neumann}.
 
\section{D-branes in 2d String Theory}
We now turn to the case of 2d string theory, for which
we need to  include the free boson $X$. For a free
boson there are standard boundary states 
\bea
\bs_N^{(X)}=exp\left(-\sum_n{1\over n}\alpha_{-n}\tilde{\alpha}_{-n}\right)|vac\rangle
\\
\bs_D^{(X)}=exp\left(\sum_n{1\over n}\alpha_{-n}\tilde{\alpha}_{-n}\right)|vac\rangle
\eea
corresponding to Neumann and
Dirichlet boundary conditions respectively.
Here $\alpha, \tilde{\alpha}$ are the left- and right-moving
modes of $X$.

We  find then the boundary states of 2d string theory by
tensoring the state found in the previous section
with either $\bs_N^{(X)}$ or $\bs_D^{(X)}$, yielding 
a D0-brane or a D-instanton respectively.

The lowest lying modes in the open
string sector are the open string tachyon
and the center-of-mass operator. 
To understand the dynamics of the D-brane,
one needs to write an effective action
for these fields.
This is more difficult than in the flat space case, because
we have defined the D-brane  as a boundary state instead of 
as a boundary condition on the open string theory.

In the boundary state formalism, one has to
solve the sewing constraints 
on open string OPEs \cite{sew}, which are
nonlinear and generally difficult to solve. In our case,
however, we can solve them  in the asymptotic weak coupling
region by
a comparison to flat space D-branes.

In our construction 
the boundary state has the same form as a
flat space D-brane, except that
we have used the primaries
of the Liouville theory rather than
the standard primaries of flat space.
Furthermore, in the weakly coupled region, 
the interactions of the Liouville primaries are
the same (to leading order) as the interactions of the free theory. 
The solution of the sewing constraints is then
the same as in flat space and the action
is the standard flat space Born-Infeld action. 

In particular, the action for the center
of mass is 
\bea
{\cal L}={1\over g_s(X_0)}\sqrt{1-\dot{X_0}^2}
\eea
where $X_0$ was defined in eqn. (\ref{bost}).
The equation of motion is that of constant acceleration
towards the strong coupling region.

In the full Liouville theory, solving the
sewing constraints requires more information.
By KPZ scaling \cite{kpz, Liouville}, the leading 
order action scales as ${1\over
\mu}$ since it is obtained from a disc diagram.
Since the mass is no longer position dependent, we
expect this to describe a static D-brane bound
to the wall.

The D-instanton would correspondingly have an action
proportional to ${1\over
\mu}$ and therefore would produce nonperturbative
effects in the spacetime theory. These 
correspond to the famous $e^{-{1\over \mu}}$ 
effects found in the matrix model \cite{shenker}.
To perform a detailed comparison, though, one
needs to find special quantities for which the
perturbative series can be summed.

Consideration of black holes in 2d string theory \cite{witten}
suggests that Liouville theory has many more
states than suggested by the perturbative spectrum. 
The entropy of black holes suggests in fact a
Hagedorn density of states \cite{tom}, whereas the perturbative
spectrum consists only of one 1+1 dimensional field
(and some special states). The success
of entropy counting in higher dimensions
\cite{sv} suggests that D-branes and
their excitations might provide the required
states.

\section{Applications}

Since linear dilaton backgrounds are ubiquitous in
string theory, the boundary state found here has many
applications. 
Most applications would require a generalization
to incorporate worldsheet supersymmetry and GSO projection. This is 
straightforward, and the resulting boundary state is
constructed in Appendix A.

It has been recently proposed that string theories 
backgrounds with a linear dilaton component are 
holographically dual to nongravitational, nonlocal
theories \cite{hol, min, lstmore, Giveon}. The earliest
 and best known example is the
near horizon limit of the NS fivebranes, which
are  holographically dual to the little string
theories in 6 dimensions with 16 supercharges \cite{lst,lst1}. Other examples
with fewer supersymmetries and in other dimensions were given
in \cite{lstmore,Giveon}.

Our construction can be applied to many such backgrounds since
they consist of a linear dilaton part and an
often solvable conformal field theory.
However, many such backgrounds have a strong coupling region
where perturbative string theory is inapplicable.
In these cases, the description of the D-brane as
a boundary state is useful only asymptotically,
in the weak coupling region.

There are however instances where the weakly coupled string
theory is applicable everywhere \cite{Giveon}. This corresponds to
a deformation of the linear dilaton background analogous to 
turning on the cosmological constant in Liouville theory. 
In such cases, our construction 
 provides an effective way of describing the 
D-brane spectrum and interactions. These D-branes correspond to
nonperturbative states
in the holographically dual theory.

In particular, some of the backgrounds in \cite{Giveon} are T-dual
to noncompact Calabi-Yau spaces with a slightly deformed singularity
\cite{tdual}. 
In such cases, the D-brane spectrum
might be more transparent in the linear dilaton description
of the background.

As mentioned already in section 4, D-branes and
D-instantons can shed light on nonperturbative
effects in 2d string theory. In particular,
counting the microscopic states of black holes
in 2d string theory requires an understanding of
the quantum mechanics of the D-branes in this background.
In addition, $e^{-1/g}$ effects in the matrix quantum mechanics
should be 
 related to
D-instanton effects in the Liouville theory.

Finally, one can speculate on a holographic relation
between the c=1 matrix model and Liouville theory,
along the lines of \cite{mald}. The world-volume theory
on a large number $N$ of D0 branes in a linear dilaton
background is a quantum mechanics of $N\times N$ matrices.
On the spacetime side, the D0 branes act as a source
for the tachyon, and hence the spacetime background 
around $N$ D0 branes is a linear dilaton background with
a nonzero cosmological constant. 
The analogue of the AdS/CFT conjecture would then state
that string theory in this background is holographically
described by the matrix quantum mechanics, which is
the c=1 matrix model/Liouville theory correspondence.
To make this precise, one needs a better understanding of the 
world-volume action of the D0-branes and the role
of the double scaling limit in the spacetime geometry.

\section {Acknowledgements}

 We thank T. Banks, I. Brunner, M. Douglas, A. Giveon, J. Harvey, D. Kutasov,
 G. Moore, H. Ooguri, J. Polchinski and  S. Shenker for useful conversations
and comments. 

Part of this work was done at
 the University of Amsterdam, and at the ITP 
in Santa Barbara. We thank both institutions for 
providing a stimulating and pleasant working environment.

This reasearch was supported in part by NSF grant No. PHY94-07194, 
and by DOE grant DE-FG02-96ER40959.

\section{Appendix}

In section 2, we reviewed a construction of a boundary
state preserving conformal invariance and based on a single 
conformal primary. In order to incorporate worldsheet supersymmetry
we need to provide a slight generalization of that
construction.

Suppose we are given isomorphic left and
right chiral algebras, and 
left and right primaries $|\Phi_L\rangle, |\Phi_R\rangle$, such that:
\be
\Omega|\Phi_L\rangle=|\Phi_R\rangle
\ee
where $\Omega$ is the isomorphism between the chiral algebras. Based
on the module of $|\Phi_L\rangle \times |\Phi_R\rangle$, one
can build a boundary state satisfying
\be
(T_\Omega-\tilde{T})\bs=0
\ee
where $T,\tilde{T}$ are any generators of the chiral algebra,
and $T_\Omega$ is the image of $T$ under the action of $\Omega$.

The construction is identical to eq (\ref{contra}).
\bea
|B\rangle\rangle = \sum_{I,J} M_{IJ}^{-1}L_{-I}\tilde{L}_{-J}|h\rangle
\eea
where here $L_{-I},\tilde{L}_{-J}$ are products of modes of
$T_\Omega,\tilde{T}$ respectively. $M_{IJ}$ is the contravariant form,
which is identical for the left and right moving algebras since
they are isomorphic.

In the case of $N=1$ supersymmetry on the worldsheet, one has
to consider the supersymmetric extension of the conformal
algebra: the NS algebra or the R-algebra. In the NSNS and the RR
sector one can apply the above construction and build a boundary state
based on each superconformal primary
\footnote{This can work in the R-NS sector only if the
NS and R algebras are isomorphic, which is the case when the theory 
is spacetime supersymmetric, and therefore has a N=2 algebra, and a spectral flow 
operator which acts as an isomorphism of the NS and R sectors.}.

The boundary states $\bs_{NSNS}$ and $\bs_{RR}$ have to 
be combined in a manner ensuring a tachyon free
open string sector. This is equivalent to imposing
a GSO projection in the open string sector. Prior to that,
one has to apply a GSO projection to the boundary state
(i.e. in the closed string channel). We now describe
how to achieve that goal. For
simplicity we concentrate on the NS-NS sector. 

The $N=1$ superconformal algebra has an automorphism
which reverses the sign of the supercharges, leaving the
bosonic generators intact. This automorphism is used to perform 
the GSO projection.

In order to construct a GSO even boundary state one
can  build boundary states $\bs_\eta$
satisfying
\bea
(L_n-\tilde{L}_{-n})\bs_\eta=0\\
(G_r-\eta\tilde{G}_{-r})\bs_\eta=0
\eea
where $\eta=\pm 1$. The action of the GSO projection on
$\bs_\eta$ is then given by
\bea
(-)^{F_L}\bs_\eta=S_L\bs_{-\eta}\\
(-)^{F_R}\bs_\eta=S_R\bs_{-\eta}
\eea
where $S_L, S_R$ are the transformation properties
of the primaries. The primary has to
be GSO even to be physical, therefore we restrict
to $S_L=S_R=1$. The GSO
invariant boundary state is then:
\be
\bs=\bs_\eta-\bs_{-\eta}
\ee

This boundary state couples then only to physical
(GSO-even) states.

\end{document}